# Spintronic nano-devices for bio-inspired computing

J. Grollier, *Member, IEEE*, D. Querlioz, *Member, IEEE*, M. D. Stiles, *Senior Member, IEEE*

**Abstract** — Bio-inspired hardware holds the promise of low-energy, intelligent and highly adaptable computing systems. Applications span from automatic classification for big data management, through unmanned vehicle control, to control for bio-medical prosthesis. However, one of the major challenges of fabricating bio-inspired hardware is building ultra-high density networks out of complex processing units interlinked by tunable connections. Nanometer-scale devices exploiting spin electronics (or spintronics) can be a key technology in this context. In particular, magnetic tunnel junctions are well suited for this purpose because of their multiple tunable functionalities. One such functionality, non-volatile memory, can provide massive embedded memory in unconventional circuits, thus escaping the von-Neumann bottleneck arising when memory and processors are located separately. Other features of spintronic devices that could be beneficial for bio-inspired computing include tunable fast non-linear dynamics, controlled stochasticity, and the ability of single devices to change functions in different operating conditions. Large networks of interacting spintronic nano-devices can have their interactions tuned to induce complex dynamics such as synchronization, chaos, soliton diffusion, phase transitions, criticality, and convergence to multiple metastable states. A number of groups have recently proposed bio-inspired architectures that include one or several types of spintronic nanodevices. In this article we show how spintronics can be used for bio-inspired computing. We review the different approaches that have been proposed, the recent advances in this direction, and the challenges towards fully integrated spintronics-CMOS (Complementary metal – oxide – semiconductor) bio-inspired hardware.

## I. Introduction

### A. Bio-inspired computing

BIO-INSPIRED, or neuromorphic, computing takes inspiration from the way the brain computes to increase the energy efficiency and computational power of our data processing systems. Biological systems have impressive computing abilities at very low power consumption levels. For example, humans are able to recognize people they barely know in just a fraction of second from a three-quarter view of their face in a crowd. Research in bio-inspired computing is driven in part by the need to invent new ways to automatically make sense of the massive amount of digital information we generate every day. Neural networks, which are extremely efficient at recognition, classification and prediction tasks, are intrinsically suited for this purpose [1] and many major companies are now investing massively in artificial intelligence research. In a recent scientific breakthrough, the machine-learning community has developed extremely efficient neural network algorithms. These deep neural networks [1] are inspired by the hierarchical structure of the cortex and are already the working principle behind the software for virtual assistants on smartphones, and for a wide range of massive classification tasks [2], [3].

Another reason for research in biological computing is to reduce the energy consumption used to perform the tasks mentioned above. The performance of processors that drive modern computing is limited by their excessive power dissipation. The amount we compute has a significant impact on global energy use. Today, information and communication technologies consume more than 5 % of the electrical energy generated in the world and this number is expected to continue growing [4]. Following current trends without rethinking the way we compute can contribute to energy shortages and environmental issues. Not only are human brains very good at tasks like recognizing faces, we do so using a million times less power than supercomputers [5], [6] do when performing these complicated tasks. The development of low-power bio-inspired computing will help address these issues.

Existing implementations of neural networks are constructed in software that runs on conventional computers rather than an attempt to imitate the efficient hardware of biological systems. Biological systems require very little power to operate for many reasons, including that their densely connected architecture allows them to compute in parallel. When mapped on the sequential architecture of existing processors, bio-inspired algorithms lose their most

J. Grollier is with Unité Mixte de Physique CNRS, Thales, Univ. Paris-Sud, Université Paris-Saclay, 91767 Palaiseau, France (e-mail: julie.grollier@thalesgroup.com).

D. Querlioz is with Centre de Nanosciences et de Nanotechnologies, Univ. Paris-Sud, CNRS, Université Paris-Saclay, 91405 Orsay, France (e-mail: damien.querlioz@u-psud.fr).

M.D. Stiles is with the Center for Nanoscale Science and Technology, National Institute of Standards and Technology, Gaithersburg, Maryland 20899-6202, USA (e-mail: mark.stiles@nist.gov).

This work was supported in part by the European Research Council ERC grant bio*SPIN*spired n°682955, the European FET-OPEN Bambi project n° 618024 and the French ANR MEMOS n° ANR-14-CE26-0021-01.

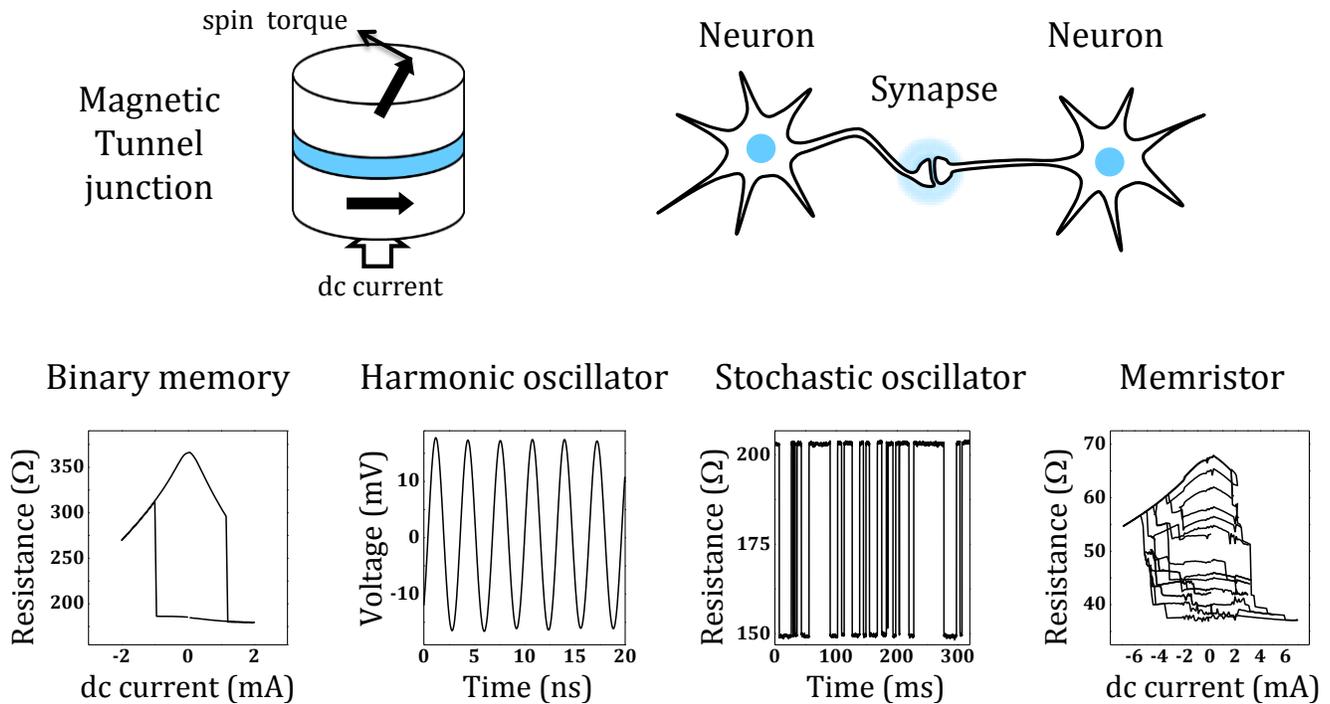

Fig. 1. Principle and multi-functionality of spin torque nanodevices. Top: a dc current injected through a magnetic tunnel junction creates a spin torque acting on the magnetization. The resulting magnetization dynamics generate resistance variations which can help mimic important functionalities of synapses and neurons. Bottom: Different types of responses can be obtained by varying the geometry of the tunnel junction and the bias conditions, such as applied field or current. Here four different responses are shown. Binary memories and memristors are interesting for emulating synapses, while harmonic and stochastic oscillators can mimic some properties of neurons or assemblies of neurons.

precious qualities: speed, defect tolerance and low energy consumption. Therefore, the most optimal solution for low-power bio-inspired computing is to fabricate networks of interconnected components to realize parallel computation on chip [7]–[10].

This vision raises two challenges. The first challenge is the scale of the network that needs to be built in order to perform interesting tasks. To appreciate the scale of these networks, the brain possesses about $10^{11}$ neurons interconnected by close to $10^{15}$ synapses, which even the world's largest supercomputer cannot simulate. Both neurons and synapses perform complex operations to allow for learning and adaptation. CMOS, as the mainstream technology today, is an excellent substrate for building such systems. However, existing CMOS devices, transistors, cannot be the entire solution. The high number of transistors required for imitating both neurons and synapses, and the related power dissipation issues limit the prospects of large scale and dense stacking [7], [11]. Existing all-CMOS-based prototypes of neuromorphic systems developed in academia (e.g. the Human Brain Flagship consortium in EU [10], [12]) and industry [13] have restricted capabilities.

A key to progress can be to invent and fabricate CMOS compatible nanodevices that will be responsible for a large part of the computation by emulating neurons and synapses directly at the nanoscale. For example, a neuromorphic chip developed by a DARPA consortium is designed so that its CMOS fixed synapses, which require offline training by a separate, conventional computer, could be replaced by matrices of tunable nano-synapses [8], which would allow the chip to learn. Toward this end, today a huge research effort tries to realize dense arrays of nanodevices called memristors on top of CMOS neurons, because a single memristor can emulate a synapse [14]–[22].

The second challenge towards building neuromorphic chips is that the existing bio-inspired computing models are abstract. They need to be rethought and adapted to be realized efficiently in hardware. Therefore, the materials, the physics that will allow nanodevices to embody interesting functions, the overall hybrid CMOS-nanodevice architecture and the bio-inspired computing models need to be developed together.

*B. Why spintronics ?*

Since the early developments of neural network theory, magnetic materials have been used as model brain-like systems. In particular, the transitions from disordered to ordered phases occurring in magnetic systems (e.g. ferromagnetic ordering at Curie temperature) are reminiscent of phase transitions observed in biological neural assemblies [23]. In 1982, John Hopfield was the first to make a direct link between neural networks and physical models [24]. He considered an Ising model, where the synaptic connections are emulated by couplings between individual spins. Since then, other models of Ising neural networks have been proposed, especially taking advantage of the many metastable states in spin glasses [25], [26]. However these models require controlling the coupling between each pair of spins for the neural network to learn. In real spin glasses, the coupling between the spins is set by the materials and geometry. It is therefore impossible to adjust, explaining why Ising neural network models have never been implemented in material systems. However, recently, models of neural networks have been developed that could be more easily transposed to hardware thanks to less stringent requirements for learning [27]–[30].

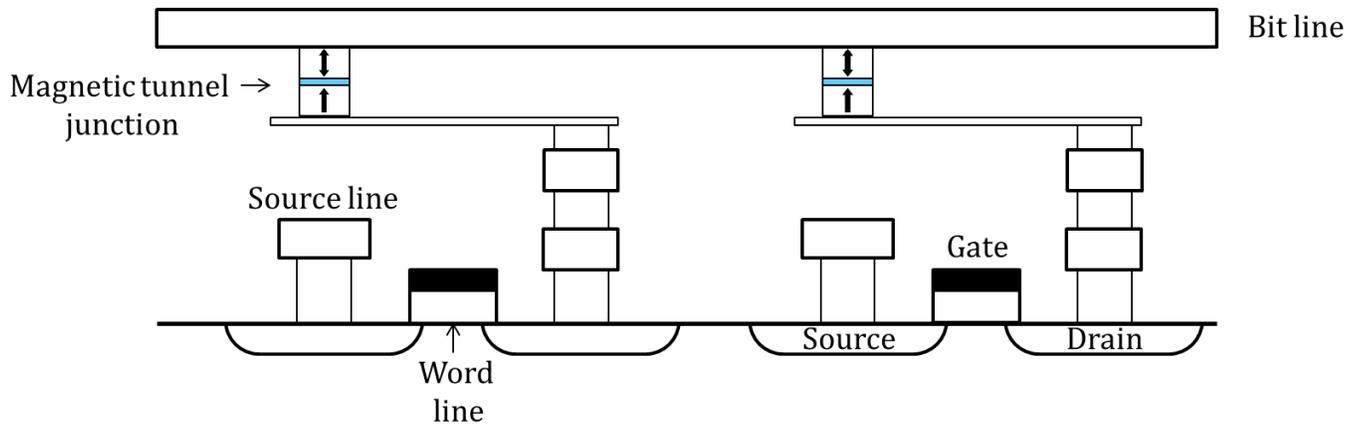

Fig. 2. Schematic view of a Spin-Torque Random Access Memory. To address a particular magnetic tunnel junction, a voltage is applied to the word line, creating a connection via the transistor below between the associated source line and all of the bit lines. A current is passed through the appropriate bit line to the selected source line to either read the state of the magnetic tunnel junction (small current) or set its magnetic state (large current). The transistors are necessary to avoid a large contribution between the selected source line and bit line through more complicated connections, so called "sneak paths."

In addition, other areas of magnetics appear to be more promising for implementations – in particular, there has been substantial progress in developing spintronic devices that could be important for bio-inspired computing. These devices are based on new ways that have been discovered for measuring the magnetic states locally, through giant magnetoresistance [31], [32] and tunnel magnetoresistance [33]–[35], and for controlling the magnetization states of nanodevices, through spin transfer torque [36], [37]. One such magnetic device, the spin valve, consists of two thin film metallic magnets separated by a non-magnetic metallic layer. All the layers are typically in the 1 nm to 10 nm thickness range. Usually, the magnetization of one of the layers is pinned by coupling it to an antiferromagnet. The magnetization of the other layer is free to respond to external stimulus. The changing relative orientation of the two magnetizations changes the resistance of the structure, i.e. the giant magnetoresistance effect, allowing electrical determination of the magnetic state of the device. In a magnetic tunnel junction, the metallic spacer layer is replaced by an insulating layer that is thin enough for electrons to tunnel between the two magnetic layers. Such a device is illustrated in Fig. 1. There, the change in the resistance of the tunnel junction with changing relative orientation of the magnetization is referred to as the tunnel magnetoresistance.

In both of these cases, the electrical resistance of the devices depends on the relative orientation of the magnetizations. This dependence can be understood in a two-current model in which the current through a ferromagnet is carried by two types of electrons, majority and minority. The resistances of the two types of electrons are different in the ferromagnet, so more current is carried by one type and the total current is said to be spin polarized. The spin polarization of the current remains largely unchanged when passing through the intermediate layer. It then interacts with the other magnetic layer, resulting in a low resistance if the properties of the layers are matched so that one type of electrons sees the lower resistance in both layers, and a higher resistance if not. For spin valves [38], the resistance can differ by 50 % between the configuration with the magnetizations parallel to each other and that with the magnetizations antiparallel. For tunnel junctions [33]–[35] the variation can be up to 600 %. The dependence of the resistance on the state of the device, which can in turn depend on its history, is a useful attribute of these devices for bio-inspired computing, as it couples the magnetic state of the nanodevice with its electrical properties.

Another useful attribute of both of these devices is that it is possible to change the state of the device by passing a current through them, through the spin transfer torque. Spin transfer torques are another consequence of the spin polarized current flowing in these devices. These spin currents carry angular momentum which interacts with the magnetization of subsequent ferromagnetic layers. This interaction is strong enough that current densities [39] as low as $10^6$ A/cm$^2$ can cause the magnetization to reverse or cause it to precess at frequencies in the gigahertz range. The magnetization dynamics induced by these spin transfer torques are converted into resistance variations due to magnetoresistive effects. In addition to the resistive readout and electrical manipulation of spin-torque nanodevices, spintronic devices possess several other virtues, which we discuss below, for bio-inspired computing [40].

**Spin transfer torque memory is close to market.** In the last few years, significant progress has been made towards the commercialization of spin transfer torque magnetic random access memory (STT-MRAM), illustrated in Fig. 2 [41].

Prototypes with 256 Mb storage have been demonstrated [42]. These results, combined with outstanding endurance and back-end-of-line CMOS compatibility, suggest that STT-MRAM is in good position to become a commercially viable non-volatile memory. Several academic and industrial teams are already taking the next step, building electronic circuits with embedded magnetic memory [43]–[46], and exploiting the physics of spin torque towards enhancing the functionality of Boolean logic circuits [47], [48]. This is important as the availability of STT-MRAM for general purpose memory will provide opportunities for developing new devices and more advanced schemes such as bio-inspired computing

**Spin transfer torque allows building a wide range of nanodevices from the same material structures.** Spin transfer torques act differently depending on the magnetization configuration, which can in turn be controlled by choosing the proper materials and geometry [40]. This flexibility may allow the implementation of different functionalities using the same materials stack but fabricating different device geometries and then changing the bias conditions during use. The functions illustrated in Fig. 1 can be particularly useful for bio-inspired computing. Binary memories [49], [50] store information. Spin-torque nano-oscillators are tiny oscillators that can generate ac voltages with frequencies larger than 50 GHz when biased with a dc current [51]. Whether harmonic or stochastic [52], they can emulate neural oscillators. Finally, the spin-torque memristor [53], [54], a tunable nano-resistor developed recently, can be used as a nano-synapse. The flexibility of spin-torque nano-neurons and nano-synapses will offer the possibility of implementing a wide range of computing concepts, and realizing reconfigurable architectures that can switch between computational modes.

**Spin-torque nanodevices are highly cyclable**. Magnetic tunnel junctions can be switched back and forth more than $10^{15}$ times without degradation [43]. In the lab, we have measured spin-torque nano-oscillators for years without their failing. This cyclability is important for implementing bio-inspired hardware that can, like the brain, reconfigure continuously to learn and process new features in an ever-changing information flow.

**Spin transfer torque driven junctions are model non-linear dynamical systems at the nanoscale.** Magnetization dynamics is non-linear, and can be tuned by adjusting the intensity of the injected current or the applied magnetic field. In particular, spin-torque nano-oscillators are non-linear frequency tunable oscillators [55]. Just like neural oscillators, spin-torque nano-oscillators can couple and synchronize due to magnetic or electric interactions [56]–[60]. This tunable non-linearity and ability to couple is a key feature for building bio-inspired computing architectures based on non-linear dynamical processes for coding, processing and storing information [61], [62]. Due to their intrinsic and tunable non-linearity, networks of inter-connected spin-torque nano-oscillators appear very suitable for implementing formal non-linear bio-inspired computing concepts.

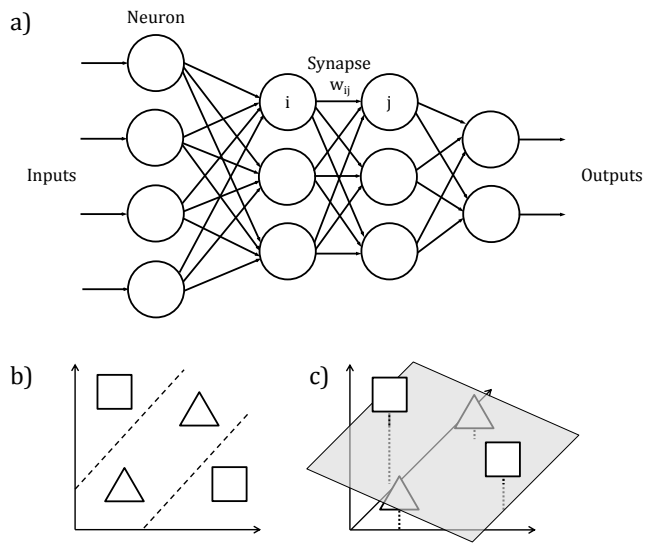

Fig. 3. Neural network. a) Four layers of neurons (circles) each take input, which they non-linearly process to produce an output signal. The output signal is passed to the next layer of neurons through the synapses (straight lines) weighted by the synaptic weight, $w_{ij}$. Signals flow from left to right. b) A simple neuron that takes the input values ($x$ and $y$ values) for different possible inputs and aims to produce an output that is different for triangles and squares. There is no linear function of the inputs that can do this separation, but non-linear functions (neurons) can. In c) a non-linear function of $x$ and $y$ produces higher output values for squares, allowing classification and reducing the information sent to the next layer.

*C. Artificial neural networks.*

Before discussing in detail how these features of spintronic nano-devices can be used for computing inspired by biology, we briefly introduce the key concepts of neural networks. Artificial neural networks are the most studied implementations of bio-inspired computing [1]. As illustrated in Fig. 3, these networks take input into layers of non-linear neurons and then pass the output of each neuron to many neurons in the next layer. In contrast to more conventional programs on computers, neural networks are not good at precise calculations. However, they excel at recognizing patterns in complex information flow, and at clustering data in an organized way. Indeed, layer after layer, the dimensionality of input data (e.g. a picture with millions of pixels) is progressively reduced, until the final output layer contains only higher-level information (e.g. dog, cat, human) [63]. The transformation of the input to relevant few outputs is possible thanks to the non-linearity of neurons. As illustrated in Figs. 3b and 3c, at each layer, the non-linearity changes the relationships between different input values. The non-linear functions of the neurons in each layer change the relationships between the inputs, making it easier to classify the inputs by associating the related ones. Associating appropriate inputs allows filtering the important features of inputs and eliminating extraneous information thereby reducing the dimension of data passed to the next layer.

The separation of inputs (e.g. finding a function that will separate triangles and squares in Fig. 3c) can be achieved thanks to the very high number of parameters that allow tuning the network response: the synaptic weights, which are the amounts by which the information transmitted from one neuron to the other is multiplied. The synaptic weights act like gradual valves for the flow of information. For classifying data, these synaptic weights have to change until the network exhibits similar behavior for similar inputs, and dissimilar response for different inputs. The rule according to which synaptic weights evolve as new inputs are presented and processed by the network is a "learning rule". In biology, the ability of synaptic weights to evolve according to neuronal activity is called plasticity.

Synaptic weights can be tuned by an external operator who knows the desired output for a given input, and who minimizes the error of the network: this is called supervised learning. One of the most efficient supervised learning rules is error backpropagation [64]. An input is presented to the first neuronal layer, propagates through the network without modifying the actual weights, and produces an output. Then, starting from the last neuronal layer the error is calculated layer by layer back to the first layer. Finally, the synaptic weights are modified by an amount proportional to the error. Recently supervised learning algorithms have shown impressive results, beating humans at image recognition [1]. They are used widely in applications such as computer vision and natural language processing. Such neural networks are very powerful within the space of data on which they have been trained. However, the training requires substantial external computer power and the networks have no way to process information that is not closely related to their training set.

Unsupervised learning occurs when synaptic weights evolve autonomously, that is without supervision, according to the local activity of the neurons connected to each synapse, similarly to what happens in the brain. In that case data clustering occurs spontaneously. The most prominent unsupervised learning rules are connected to biological models and can often be classified among "Hebbian" learning rules. The underlying principle is that "cells who fire together wire together". In other words, a synaptic weight is modified in proportion to the activity of its pre- and post- neuron [65]. Unsupervised learning methods can solve efficiently medium-size problems such as visual feature extraction [66]. The next challenge in artificial intelligence is large scale unsupervised learning. This capability allows neural networks to learn how to treat data that no operator has formerly classified or identified.

To summarize, the common features to all neural network models are: non-linearity, a high number of tunable parameters for learning and enough reproducibility in the response of the network to distinguish between different classes of inputs. These are the features that need to be created in spintronics neural networks. Section II presents preliminary approaches to implement some of these ideas. Section II.A describes the utility of magnetic tunnel junctions used as MRAM cells to fuse memory and processing in one region of space to capture the colocation of both in the brain. Section II.B discusses proposals to use magnetic tunnel junctions in the opposite limit, in which they are thermally unstable rather than being stable for ten years. In this limit, they require much less power to use. Section II.C describes how to use magnetic domain walls to implement a variety of features of both neurons and synapses. Section II.D presents proposals to take advantage of the non-linear dynamics in spintronic devices. All of these approaches face serious challenges, which are presented in Section III and a summary of this paper is given in Section IV.

## II. IMPLEMENTATIONS OF BIO-INSPIRED HARDWARE USING SPINTRONICS

### A. Fusing memory and computing

It is instructive to contrast the large scale design of traditional von Neumann computers with our brains. Computers are sequential; they are designed around a powerful processing unit that has access to all of the information stored in the computer. While many things happen at the same time in computers, all of these activities are focused on the computer doing one logical step at a time. Much of this activity is dedicated to bringing information from memory to the processing unit because memory and data processing are spatially separated. Data is continuously transported back and forth, consuming power. The communication bus between computing and memory is often called the "von Neumann bottleneck". Despite efforts toward increasing parallelism in computers, this separation of memory and processing remains a fundamental principle of traditional computers.

On the other hand, our brain functions with completely embedded processing and memory. The processing units, the neurons, are taking many inputs and producing a simple output. The neurons all work simultaneously in parallel but operate on the basis of very limited amounts of stored information, provided through the weights of the synapses which connect them to other neurons. This entanglement of memory and processing along with parallel processing are two reasons the brain is low power and fast at certain tasks.

Let us consider the simple feed-forward artificial neural network -- a canonical example of neural network -- illustrated in Fig. 3a. The synapses represent weights and are stored as floating point real numbers. When a conventional processor is used to evaluate the output of such a neural network, the computer needs to compute the state of each neuron, which is not a particularly computationally expensive task. However, to do so, the processor needs to retrieve from memory the synaptic weights of all the synapses connected to the neuron. This kind of task, which requires little computing but substantial memory access, is especially unfavorable for computers because of the separation between computing and memory. The inefficiency of bio-inspired and cognitive models on traditional computers is widely accepted [67], therefore it is attractive to design computing structures for such assignments that would fuse computing and memory [7], [11]. From a design perspective fusing computing and memory is a difficult challenge. In recent years, however, there has been considerable progress in one direction: neuromorphic chips implement neural networks with memory blocks embedded at the core of computing, [8], [10]. Currently, such chips use static random access memory (SRAM), a very fast form of memory, but one that uses substantial active and passive power and occupies a large area in the circuit [8], [17].

Therefore, these systems possess limited memory capacity. Replacing SRAM by magnetic memory could therefore dramatically improve the capability of current neuromorphic chips. Additionally, unlike SRAM, magnetic memories are non-volatile. Not only does this minimize passive power consumption but in addition, the system could be turned OFF and ON and function instantly, an especially attractive feature for embedded applications.

Therefore, the most straightforward application of spintronics within a bio-inspired system is as embedded memory to store the parameters of the system, such as the synaptic weights in the case of a neural network. This prospect is near, as it has been technologically demonstrated that such cells can be embedded at the core of CMOS [68].

As magnetic memory becomes readily available, bio-inspired digital systems specifically designed for magnetic tunnel junction cells can also be imagined. Such systems associate small computing with memory blocks distributed all over computing blocks. A first digital bio-inspired system with magnetic tunnel junctions working along this idea has already been demonstrated [69]. This associative memory achieves 89 % energy reduction in comparison to approaches using conventional hardware. One can also imagine going further and entirely fusing magnetic tunnel junctions with logic, therefore not having any difference between logic and memory blocks. It is for example possible to design logic blocks where some inputs are memorized parameters stored in magnetic tunnel junctions [70], [71]. Such logic gates might give rise to systems which entirely eliminate all the energy and delays associated with memory access, and that would probably be well adapted to bio-inspired models. However, their design brings considerable challenges and their potential has not been fully achieved.

*B. Leveraging noise for computing*

It is also possible to use magnetic tunnel junctions for different purposes than non-volatile memory cells. MRAM is designed to be thermally stable so that information is preserved for ten years. Therefore, the energy consumption required to switch perfectly non-volatile magnetic tunnel junctions is relatively high, typically 100 fJ [72], as compared with 23 fJ per synaptic event (considering that there are in average ten thousand synapses per neuron in the brain) [73]. In addition, magnetic random access memory cells are required to have a minimum variability, which imposes severe constraints on nanofabrication. If MRAM cells are predominantly used passively, this stability is advantageous because the passive power use is zero. On the other hand, writing new information in the MRAM cell requires energies much higher than thermal. If the circuit requires frequent changes in the stored information, MRAM is not particularly low power [43], [68].

In the opposite limit, in which the barrier between the two states is comparable to the thermal energy, changing the state of the tunnel junction requires much less power. Neuroscience data indicates that the brain and its components operate at this thermal limit making them very noisy [74]. Neurons and synapses consume very little energy but are unreliable and display stochastic behavior [75]. Nevertheless computations in the brain are reliable [76]. One common interpretation of this property is that the brain compensates for the high noise and variability of its individual components by redundancy [77]. Apparently, the brain finds the optimum tradeoff between lowering the energy and the reliability of the computation to be very close to the thermal limit. In spintronics, a similar strategy is conceivable when the barrier between states in a magnetic tunnel junction is significantly lowered as compared to MRAM. We can imagine lowering the usual criteria used for designing magnetic memories when designing magnetic nanodevices for bio-inspired computing. By allowing noise, variability and stochasticity, the energy consumption of magnetic nano-objects can be lowered, and their size can be reduced below 20 nm. Additionally, embracing such behaviors can allow spintronics devices to display richer, more complex physics, and therefore make them more analogous to biology's nanodevices. For example, unlike in many models of artificial neural networks, biological synapses do not only act as a real number weights: they have rich dynamics and behaviors, which are harnessed by the brain for computing. As biology exploits the rich physics of its synapses for computation [78], one can use the dynamics of spin transfer torque switching physics for computation. This general idea of harnessing device physics for bio-inspired computation was pioneered by Carver Mead in the late 1980's [79]. He proposed using transistors in weak inversion to implement neural network blocks, an approach which is still used in large neuromorphic systems [7], [9].

In the following, we describe a few ideas on how to compute with stochastic magnetic nanodevices.

**Probabilistic magnetization switching**. Switching of magnetic devices is intrinsically probabilistic due to the importance of thermal effects [80], [81]. When a magnetic field or spin transfer torque is applied to a magnetic tunnel junction, it creates a probability rate for switching. For memory applications, the amplitude and durations of current pulses applied for switching are chosen so that the probabilistic effects result in an acceptable error rate [80], [82]. If the currents or pulse durations are reduced, it is possible to tune the switching probability to any chosen value. If successive switching events do not follow each other too closely (with a frequency smaller than a few hundred megahertz), the switching probabilities are not correlated. By setting the switching probability close to 50 %, spin transfer torque has been used to generate true random numbers, using limited post-processing [83].

It is also possible to harness these probabilistic effects: magnetic tunnel junctions can be considered as a form of memory with "stochastic programming", when used with short, low-energy voltage or current pulses. Such a memory is reminiscent of some models in computational neuroscience or in machine learning, where synapses do not feature floating point real number weights, but binary weights programmed stochastically [84]–[86]. In particular, spin-torque driven magnetic tunnel junctions can implement a stochastic version of spike timing dependent plasticity [87].

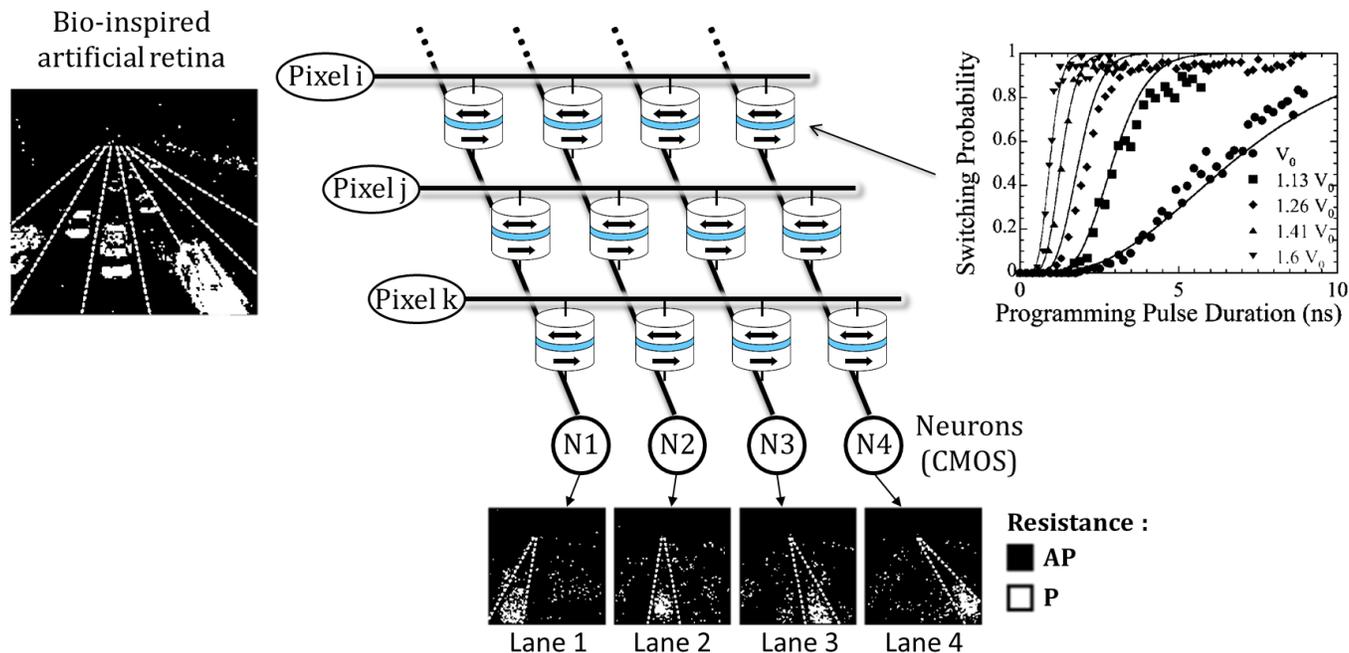

Fig. 4. Simulations of learning through probabilistic switching of magnetic tunnel junctions [82]. The synaptic crossbar array (center schematic) consists of magnetic tunnel junctions for which the probability to switch depends on the programming pulse duration and amplitude (right graph). Here, for learning, pulses are chosen so that junctions have only a slight probability to switch. Input pulses code for each pixel amplitude in a video of cars on a highway taken with a bio-inspired artificial retina (left image). Output pulses are generated by the output neurons $N_i$ if the input pulses weighted by the junctions' conductances in each column exceed a threshold. The switching of junctions depending on input and output pulses evolves according to spike timing dependent plasticity. The junctions' states are initially random but after the input video has run for some time, the weights stabilize to a configuration such that each output neuron specializes to recognize cars in each lane of the highway (images at the bottom). In other words, the neural network made of stochastic magnetic tunnel junctions has autonomously learnt to count cars in each lane.

Spike timing dependent plasticity (STDP) is an Hebbian learning rule inspired by biological measurements [88], [89]. Even though the synapse transmits information in one direction, it is influenced by the firing of both the pre- and post-synaptic neurons. If they spike together in a short time window, the synaptic weight is modified. It increases if the post-neuron fires after the pre-neuron, indicating a causal relation, and decreases otherwise. It has been shown recently that memristor nanodevices can implement spike timing dependent plasticity [15], [16], [19]. By carefully choosing the shape of neuronal voltage pulses, their resistance can evolve autonomously and gradually according to the firing of pre-and post-neurons [90]. Simulations indicate that unsupervised classification of features in input data flow is possible in systems where different neural layers are connected by memristor crossbar arrays [91].

In binary devices such as magnetic tunnel junctions, the resistance cannot evolve gradually according to the pre-and post-neurons activities, but it can evolve probabilistically. The probability of a junction switching during a voltage pulse can be tuned between 0 and 100 % through the amplitude of the pulse. This allows implementing a probabilistic STDP learning rule, where the relative timing between neural spikes does not determine the amplitude of an analog synaptic weight modification, but the probability to switch a binary weight. How this works can be understood as follows. When a neural network learns, it is essential that each learning event changes the network only slightly. The canonical method to achieve this is to have synapses with real number weights that are updated only slightly at each learning step. An alternate method is to use binary synapses, which have only a slight probability to change at each learning step. Using discrete synapses makes learning slower but endows the network with an increased memory stability [85], [86].

Recent simulations [92] explore the capability of magnetic tunnel junctions for stochastic spike timing dependent plasticity. It shows that a system equipped with magnetic tunnel junctions implementing a highly abstracted form of stochastic STDP can learn complex tasks such as detecting cars in a video (Fig. 4). Interestingly, the system is robust to device variations: due to device mismatch, each magnetic tunnel junction has a different switching probability, but this can be tolerated to a wide extent by neural networks (Fig. 4). It should also be noted that it is possible to combine stochastic synapses to recreate analogues to multilevel synapses. This is necessary for a neural network to accomplish hard tasks such as image recognition [11], [93].

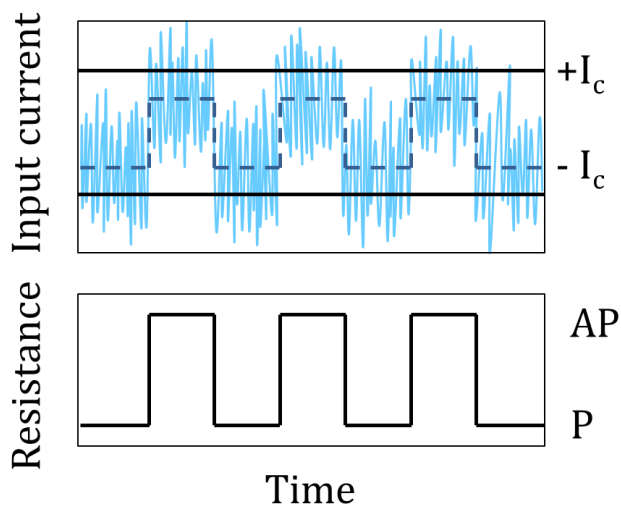

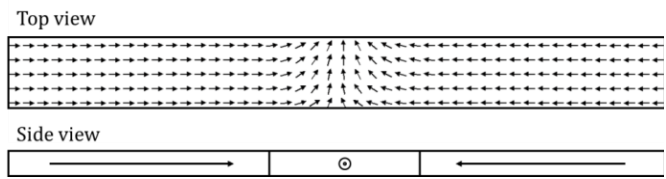

Fig. 6. Magnetic domain wall. The arrows indicate the direction of the magnetization. For typical thin films, the energy is lower when the magnetization is parallel to the side of the structure, so in thin film wires, it tends to lie in the plane along the wire. There are two possible directions for domains. Where they meet is a domain wall, where the magnetization rotates continuously from one direction to the other.

Fig. 5. Principle of stochastic resonance applied to magnetic tunnel junctions. The dashed curve shows the input signal, which does not reach the thresholds for switching (heavy solid curves labeled $+I_c$ and $–I_c$). When an appropriate level of noise is added (solid curve) the current does cross the critical currents and the device switches. Even though the noise fluctuates below the critical current, the device stays in the desired state because the current never crosses the threshold for switching in the other direction. The bottom panel gives the resistance of the device due to the switching caused by the noise plus the signal. The resistance closely matches the input signal.

**Stochastic resonance.** A common method deployed by biological organisms to exploit noise for computing is stochastic resonance [94]. The principle is illustrated in Fig. 5. Consider a dynamical system that can compute if the input signal reaches a given threshold. In the absence of noise, if the excitation signal is weaker than the threshold, the sensor is unable to detect the small input. However, in the presence of noise, the signal will be amplified by fluctuations at its maxima, and thus able to trigger the detection. Stochastic resonance is widespread in nature, and has been observed in various biological systems, such as the behavior of feeding paddlefish [95], neural models [96] and many others.

Magnetic tunnel junctions, which are typical double well systems with a threshold (the critical current for switching), exhibit stochastic resonance [97]. Some applications in electronics, especially for audio (dither) processing make use of stochastic resonance by adding noise to the system. For audio processing the noise has to be added specifically for this purpose because current electronic circuits are designed to eliminate all noise sources. However, in a bio-inspired computing context, noise is omnipresent, and stochastic resonance does not require additional noise sources [98], [99]. We can therefore envisage constructing spintronic circuits harnessing stochastic resonance for bio-inspired applications, taking inspiration for example from cochlear implants [100].

### C. Propagating magnetic information in devices and arrays

In the brain, efficient information propagation is vital [101]. Neuroscientists have observed that many neurological disorders are due to connectivity issues between spatially distributed brain regions [102]. In spintronics, information can be represented in different ways. It can be a magnetization state or texture, an electric current or even a spin current. In the following we show how the propagation of magnetic information can be used in individual magnetic nanostructures to capture important brain-like functions and as a principle for computing in arrays of interacting magnetic nanodevices.

**Magnetic domain walls for multilevel memristive magnetic synapses.** The strength of the coupling with which synapses transmit information between the neurons they connect depends on the past activity of those neurons. The efficiency evolves continuously and gradually based on the electrical impulses from those neurons, a property called plasticity. Plasticity allows neural networks to learn and reconfigure. Magnetic devices are particularly well adapted for implementing such plasticity [103], [104] due to their memory effects and tunability. In particular, leveraging magnetic domain wall displacement in a magneto-resistive structure, in contrast to switching a magnetization in one shot and uniformly, can be used to implement synaptic plasticity.

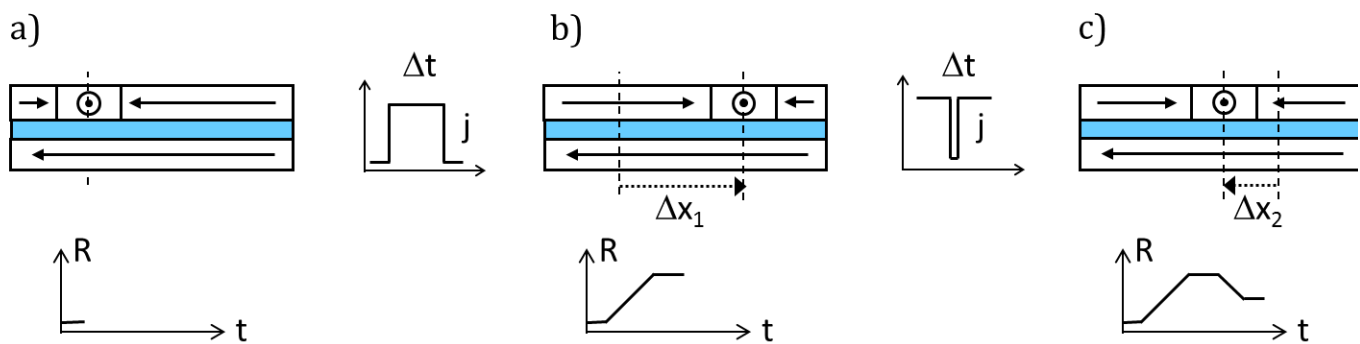

Fig. 7. Principle of the spintronic memristor based on magnetic domain wall motion. The position x of a domain wall in a magnetic trilayer determines the fraction of parallel and antiparallel domains and sets the resistance of the junction. When a current pulse is injected, the domain wall is expected to move by a quantity $\Delta x$ proportional to the pulse duration and amplitude, in other words, to the charge. In addition, the direction of the domain wall motion is set by the sign of injected current. The trilayer resistance depends on the charge that was previously injected, making it a memristor device.

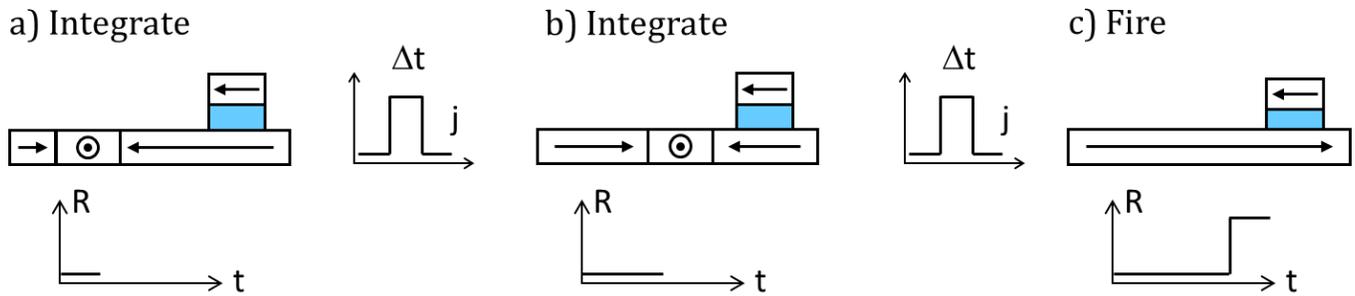

Fig. 8. Neural integration based on magnetic domain wall motion [104]. A domain wall is initially positioned at the end of a magnetic stripe further away from the magnetic tunnel junctions. After each pulse injected in the magnetic stripe, the domain wall moves towards the junctions by a given amount. During the integration phase a) and b), the motion of the magnetic domain wall does not modify the junction resistance. When the domain wall passes below the junction, the magnetization configuration of the junction switches from parallel to antiparallel, its resistance jumps to the high state: this is the firing phase c). After firing, the configuration has to be reset to a).

As shown in Fig. 6, a magnetic domain wall is a magnetic object separating regions with uniform magnetization. Magnetic domain walls are easily created in magnetic structures with a stripe shape. They can then be displaced by spin-torque through the injection of an electrical current either in the stripe or perpendicularly to its plane [105]. In an ideal stripe, a current pulse of amplitude $I$ and duration $\Delta t$ displaces a domain wall by a distance $\Delta x$ proportional to $I\Delta t$, in other words, proportional to the amount of charge $\Delta q$ that has been injected [106]. As illustrated in Fig. 7, when this stripe is used as one of the layers of a spin-valve or a magnetic tunnel junction, current pulses give gradual displacements of a domain wall, resulting in turn in gradual variation of resistance $\Delta R$, such that $\Delta R$ is proportional to $\Delta q$ as well [53], [54], [107]. This dependence of resistance on the charge is the hallmark of memristor devices. Such memristive behavior has been demonstrated in magnetic tunnel junctions with more than 15 intermediate resistance states [108]. Recently, it has also been shown that similar smooth magnetization variations can be triggered by spin-orbit torques in a magnetic stripe on top of an antiferromagnetic layer [109]. Memristive-like features can then be obtained by fabricating a tunnel junction on top of the bilayer stripe.

Such spintronic memristors may be used as multilevel synapses, similarly to many schemes proposed for other memristive technologies [15], [20], [110], [111]. In such proposals, the conductances of the memristive devices act as synaptic weights: inputs are presented as voltages, which are converted into weighted currents by the nanodevices. They can be naturally coupled to either CMOS neurons [112] or spintronic neurons as described in the next section. As we have seen previously, oxide memristors allow an easy implementation of the spike timing dependent plasticity learning rule, potentially leading to neural networks learning autonomously. Learning through spike timing dependent plasticity has not yet been demonstrated in spintronic memristors.

**Magnetic domain walls for neural integration**. In the brain, neurons integrate the sum of the weighted synaptic currents they receive. When the total integrated input current exceeds a threshold the neuron emits a spike and resets. This behavior is called "integrate and fire." Both the integration phase and the non-linearity associated with the threshold play an important role in neural computation.

Spintronic devices can realize neural-like integration and thresholding. Integration can be realized as described above for devices based on moving domain walls. Thresholding can be realized using a standard magnetic tunnel junction, which switches only if the amount of current it received is above the critical current. The switch of the junction resistance from ON to OFF state emulates neural spiking. After each switch the junction has to be reset to the ON state by a current pulse of opposite polarity. To realize the integration and thresholding in the same device, a tunnel junction with a bottom magnetic electrode extending as a long stripe on one side can be used, as can be seen in Fig. 8 [113]. The weighted input current to the neuron is injected in the stripe and used to move a magnetic domain wall. To illustrate the principle, let us consider that the domain wall is initially at the end of the stripe the farthest away from the junction. As information flows in the stripe as a function of the electrical activity of pre-neurons, the domain wall will gradually move along the stripe, getting closer and closer to the junction. This motion has no effect (integration phase), until the domain wall reaches the junction, and passes below it, thus switching the magnetic configuration (firing phase). Then, the device is reset and the process repeats.

In a more futuristic vision, such neurons could also operate with pure spin currents. Several theoretical works have investigated this possibility for perceptrons, which are single layer neural networks [113]. Due to the limited spin diffusion length of magnetic metals, such a scheme could only be used for small structures: conversion to charge current is necessary to connect to a network over larger distances. Optimistic assumptions on spin devices suggest that this approach could reduce power consumption very significantly with regards to charge-current based approaches [113]. Many variations of this idea are possible [114]. In particular, it could be possible to implement convolutional neural networks, a basic element of deep neural networks [115]. Of course, the success of these proposals is dependent on the success of all spin logic, which still has many challenges [116].

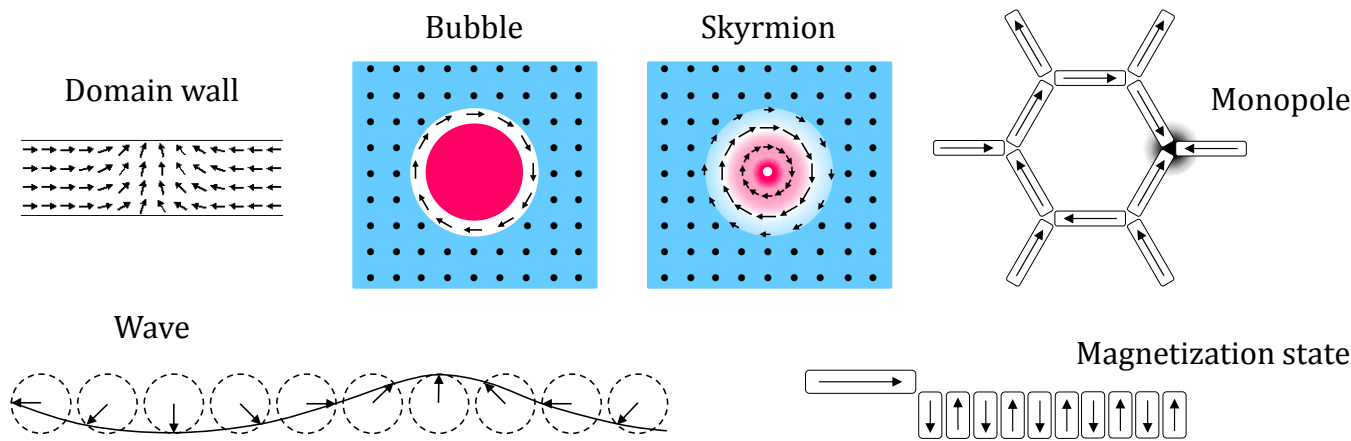

Fig. 9. Different magnetic solitons seen from a top view. Arrows are larger when they are in plane. The background color reflects the local out-of-plane component of magnetization. Domain walls, bubbles, skyrmions, and waves are all solitons in continuous media. On the other hand, the monopole is a point of frustrated interactions between bar magnets in an artificially fabricated lattice, frequently referred to as an artificial spin ice lattice. The magnetization state is one of two configurations found in an all-spin-logic device.

**Soliton propagation in arrays of interacting magnetic nano-objects.** Magnetic domain walls are not the only objects that can be displaced inside magnetic layers by currents and magnetic fields. Magnetic bubbles [117] and skyrmions [118], monopoles [119], waves [120] or even the local orientation of magnetization [121] can propagate in a controlled way (Fig. 9). It is conceivable to use these tiny solitons, rather than just charge, as the units of information in spintronic neural networks. This approach is feasible though challenging. Shift registers based on the motion of solitons have been realized, such as the magnetic bubble memory [122], or are currently investigated in industry such as the racetrack memory based on domain walls [123]. Solitons can be propagated in large arrays of nanomagnets in the framework of nanomagnetic logic [121] or spin ice [124], [125]. To realize bio-inspired computing, the challenge will be to tune these networks, so that when solitons representing an input are injected into the network, they propagate in a way that will be characteristic of this input, and easily detectable, allowing for pattern recognition and classification. Such specific cascades of events in response to specific inputs can take different forms, such as phase changes or avalanches in networks close to criticality [23], features that have been observed in the brain [126].

### D. Non-linear dynamics at the nanoscale

A whole class of computing models takes inspiration from the dynamical nature of the brain when processing cognitive data [78], [127]. Neurons and synapses are dynamical objects. Synapses evolve in time, particularly the degree to which they transmit information. The connections are decreased or reinforced according to the activity of neurons, a process which allows the network to learn. Groups of neurons can be modelled as nonlinear oscillators that adjust their rhythms depending on incoming signals [128]. The brain itself displays a wealth of phenomena characteristic of non-linear dynamical systems: synchronization of oscillating neural assemblies [129], complex transients [130], and even chaotic behavior [131].

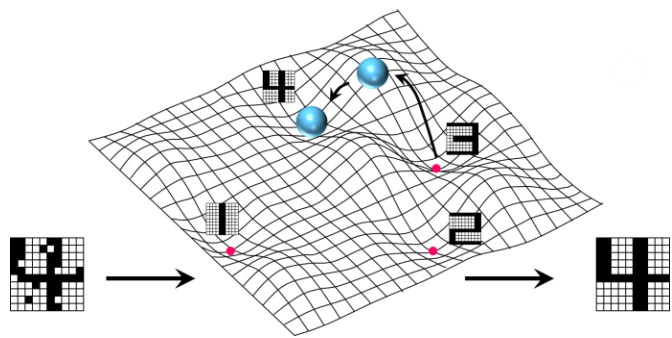

Fig. 10. Principle of Hopfield networks. Hopfield networks are distinct from networks with synapses that transmit information in one direction in that they have symmetric connections between pairs of neurons. With these symmetric connections, it is possible to define an energy of the system when the state of the system is mapped onto a position. When the system is trained to recognize particular patterns, like the four on the right, the energy of that state is a local minimum. That means that when something close to a four, like the pattern on the left, is presented to the network, it relaxes to the closest local minimum, which is the four on the right.

Neural networks with feedback, in contrast to the strictly feed-forward networks illustrated in Fig. 3a, are called recurrent neural networks. They have significant computing capabilities and can implement any kind of dynamics (fixed points, limit cycles, and chaos) [132]. Attractors in such systems can store memories. Transient dynamics can be used to process input time sequences provided by sensors or to generate trajectories as outputs for motor control [133]. Spin-torque nanodevices, which are multi-functional and tunable nonlinear dynamical nano-components, are interesting building blocks for implementing recurrent neural network models in hardware. They can be assembled and coupled in large networks in order to generate complex non-linear dynamics that imitate interesting behaviors of populations of neurons and synapses.

A well-known example of a recurrent neural network is a Hopfield network. When synapses are symmetric, that is, when information flows between each pair of neurons at the same rate in both directions, Hopfield has shown that the dynamics of recurrent neural networks derives from an energy function [24]. A network containing a large number of neurons and synaptic connections can have numerous energy minima. The energy minima correspond to dynamical attractors, which can be used to store information. As illustrated in Fig. 10, when a noisy input is presented to the system, in spite of the noise, it is in the basin of attraction of the pattern to be recognized and dynamically converges to the attractor performing a 'recognition' step.

The attractors in Hopfield networks were originally considered to be static fixed points. Following this idea, it has been recently demonstrated experimentally that arrays of coupled nanomagnets can perform pattern recognition in images by minimizing their global energy [134]. The attractors can also be the different synchronized states of networks of coupled oscillators. In 1998, Aonishi theoretically proved that a network of coupled phase oscillators with individually adjustable coupling strengths can recognize binary pattern vectors from a set of memorized patterns [135]. Most current work on bio-inspired computing with oscillators continues to be theoretical. The only existing electronic implementation, which is very recent, involves a circuit board with 8 lumped oscillators that gives a proof-of-concept without prospects for scaling up the system [30].

The dearth of hardware prototypes follows from the stringent requirements on the oscillators. In order to build a bio-inspired memory based on the associative operations of the brain, it is necessary to implement a network of oscillators that can be synchronized and in which the coupling between individual oscillators is tunable. In addition, maximizing the storage density and the efficiency of the network requires shrinking the oscillators to nanometer-scale dimensions. In this context, the nanometer size, tunability and ability to synchronize of spin-torque nano-oscillators could be disruptive. There are several proposals for interconnecting such oscillators for computing [136]–[138]. We expect that an experimental demonstration will follow soon.

The challenges for real scale applications will be to realize large networks of synchronized oscillators, to tune the couplings between oscillators, to efficiently detect the emerging synchronization patterns, and to minimize the energy consumption. Spintronics offers many approaches for tuning the coupling between magnetic oscillators needed to generate the desired synchronization patterns. When the coupling is electrical, memristors can be inserted in the current lines connecting the oscillators. When the coupling is induced by spin waves, it can be modified by spin-orbit torque locally damping or enhancing the wave amplitude.

Two approaches can be used for reducing the energy dissipation during computation. The first is to use spin-torque oscillators with a high frequency in the range of several tens of gigahertz. In this case, the computation time, given by the time to reach synchronization [139] after the initial perturbation of the network by the input, will be short, typically a few nanoseconds ($\approx$2 ns at 50 GHz), reducing the total energy correspondingly. The other solution is the opposite – to use ultra-slow, but stochastic magnetic oscillators [52], [140], [141]. For example, neural oscillators can be emulated by superparamagnetic tunnel junctions, which fluctuate randomly between the ON and OFF resistance. Instead of functioning as unstable bits, superparamagnetic tunnel junctions can be treated as stochastic oscillators that do not need any source of energy to oscillate other than thermal noise. In addition, spin-torque is particularly efficient in these junctions since the energy barrier between the magnetization configurations is small. Due to these properties, superparamagnetic tunnel junctions can be phase-locked to a weak periodic excitation [52], [142], opening the path to low power synchronization of magnetic oscillator networks.

## III. THE CHALLENGES OF SPINTRONICS FOR BIO-INSPIRED COMPUTING

**Designing modular magnetic neural networks**. Magnetic tunnel junctions are nanoresistors, as are most memory cells in other emerging technologies, such as resistive random access memories [17], phase change memories [18] or ferroelectric memories [143] etc. The main advantage of spintronics compared to other resistive memories for neuromorphic computing is the possibility to induce complex and tunable resistance dynamics through spin-torque. Like other memory cells, they can switch between fixed states allowing them to emulate synapses. But the resistance of a magnetic tunnel junction can also oscillate, spike, or show chaotic dynamics [144]. These dynamical behaviors potentially allow tunnel junctions to implement neurons at the nanoscale, a role which is not possible with other memristor technologies that require the addition of capacitors or inductors to oscillate [145].

A drawback of spintronics is that magnetic tunnel junctions have small resistance variations compared to other memory cells, with OFF/ON ratios typically equal to 2 or 3. Therefore it will not be possible to create large arrays of electrically interconnected junctions without selector devices placed under each because otherwise so-called sneak paths dominate the array [146]. In addition, fast electrical signals damp out quickly in large resistive arrays. One way to create larger networks of interacting elements could be to use magnetic coupling through dipolar fields between nanomagnets, as in artificial spin ices and nanomagnet logic arrays [119], [121]. But in any case, an organization in small modular arrays, interconnected through CMOS interfaces, will be necessary. Magnetic neuromorphic computers will require radically new architectures, using special design rules to assemble elements or devices into smaller scale circuits and then integrating such circuits into higher order operational units. Computing with ensembles of smaller neural networks follows closely the modular and hierarchical organization of the brain. Such models (deep and modular neural networks) already exist [147], and adapting them to magnetic systems will be an important challenge.

**Giving spintronic networks useful features.** Aspects of brain behavior that these circuits may inherit include spiked input and output, stochastic behavior, strong feedback, non-linearity, and operation close to the thermal limit. As we have outlined in this review, many different paths can be explored for this purpose. While most neural network models are very tolerant to variability between components (i.e. different behaviors for different neurons and synapses), the quality of computation degrades rapidly when the behavior of individual components' behavior is not consistent with itself. Therefore, generating reproducible responses in these networks will be crucial, independent of the computing substrate: domain walls, skyrmions, waves, electrical oscillations. Designing the magnetic network architectures and functionality will require interdisciplinary studies, and the development of adapted fast numerical simulation tools.

**Tuning for learning.** Once a network has been endowed with the desired function it has to be trained to give different responses to the different kind of inputs that should be differentiated. In many models, training requires being able to tune the interactions between each pair of neurons. It will therefore be a huge technical challenge to find efficient ways to tune interactions inside large assemblies of magnetic nano-objects. Here spintronics has some advantages, as many possibilities are available for tuning the information propagation between magnetic nano-objects, for example via local spin transfer torques or spin-orbit torques, electric field induced anisotropy modifications, or magnetic fields generated through close-by wires.

**Measuring the response of magnetic neural networks.** Clearly, one of the requirements for spintronics-based bio-inspired computing will be to design and use magnetic nano-devices with easily measurable states (whether they are the resistances of a junction, domain wall positions, magnetic configurations…). In any case, the standard tools used to characterize existing circuits will not work for circuits with these properties because the circuits will be inherently stochastic and will likely involve feedback. Therefore, the output will not be a simple function of the instantaneous input. To progress towards spintronic neuromorphic computing, it will be necessary to develop the measurement techniques needed to characterize such circuits. These measurements will provide feedback to research aimed at optimizing individual devices and to research on developing architectures to combine such circuits into to functioning computers. Modelling will facilitate this feedback. Thus, it is essential to bridge the device-circuit and circuit-architecture gaps by characterizing the behavior circuits of novel devices assembled and developing models of the behaviors of such circuits for use in architectures.

IV. CONCLUSIONS AND PERSPECTIVES

Neural network algorithms are already in widespread use. The next step is to realize low power computing by building chips whose organization is inspired by the brain's architecture. One of the challenges is the almost infinite number of possibilities. Undoubtedly, CMOS devices will play an important role, but it is likely that novel nanodevices will complement them by bringing important functionalities such as memory and intrinsic forms of plasticity. In this review, we have described how spintronic devices might play an important role. Magnetic tunnel junctions can bring non-volatile memory close to CMOS. In addition, magnetic nanodevices display a wide variety of behaviors that capture some of the properties of both neurons and synapses. They have the great advantage over other prospective devices in that there is already significant experience in integrating them into CMOS circuits.

To date, most ideas have not reached the experimental level, and in most cases the experiments are preliminary, making this promising field wide open for more experiments and additional ideas. Further progress will require a broad and interdisciplinary approach. Original physics should be developed to confer interesting functionalities for computing to magnetic nanodevices and magnetic circuits. At the device level, much is known about optimizing magnetic tunnel junctions that require long term stability. Not nearly as much is known about optimizing these tunnel junctions when designed to function with lower thermal stability and energy cost. Devices based on magnetic domain wall motion or other magnetic solitons are still in their infancy. While there have been demonstrations of coupling several magnetic nanodevices together, it is still not clear how to connect large numbers of devices together and even less how to compute with these assemblies. Moving from a few coupled devices to circuits of millions of neuron-like devices connection by hundreds of millions of synapses will require a number of breakthroughs in circuit design, circuit measurement, and modelling.

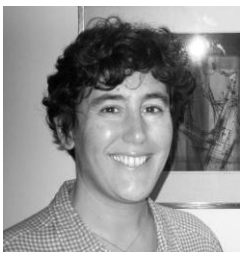
**Julie Grollier** is a researcher director in the CNRS/Thales lab in France. Her Ph.D was dedicated to the study of a new effect in spintronics: the spin transfer torque. After two years of post-doc, first in Groningen University then in Institut d'Electronique Fondamentale, she joined CNRS in 2005. Julie is Fellow of the American Physical Society, was awarded the Jacques Herbrand prize of the French Academy of Science and she is the recipient of two European Research Council grants. Her current research interests include spintronics (dynamics of nanomagnets under spin torque), and new nanodevices for cognitive computing. She is also chair of the interdisciplinary research network GDR BioComp coordinating national French efforts to progress towards the hardware realization of bio-inspired systems.

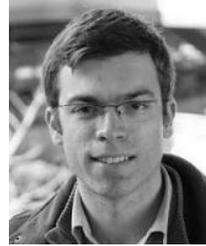
**Damien Querlioz** received the M. S. degree from Ecole Normale Superieure, Paris in 2005 and the Ph.D. degree from the Univ. Paris-Sud, France, in 2008. After postdocs at Stanford University and at CEA LIST, he became a CNRS research scientist with Univ. Paris-Sud in 2010. He develops new concepts in nanoelectronics and spintronics relying on bio-inspiration. His research interests also include the physics of advanced nanodevices. He leads the ANR CogniSpin project, which investigates the use of magnetic memory as synapses. He leads the CNRS/MI DEFIBAYES project and is a one of the lead PI of the FP7 FETOPEN BAMBI project, which explores the new paradigms for nanolectronics based on Bayesian inference.

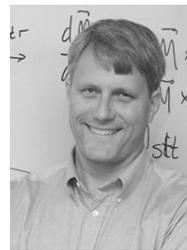
**Mark D. Stiles** is a NIST Fellow in the Center for Nanoscale Science and Technology at the National Institute of Standards and Technology, USA. He received a Ph.D. in Physics from Cornell University followed by postdoctoral research at AT&T Bell Laboratories. Mark's research at NIST has focused on the development of theoretical methods for predicting the properties of magnetic nanostructures. He served the American Physical Society as the Chair of the Topical Group on Magnetism and on the Executive Committee of the Division of Condensed Matter Physics. He was a Divisional Associate Editor for Physical Review Letters and is currently on the Editorial Board of Physical Review Applied. Mark is a Fellow of the American Physical Society, and has been awarded the Silver Medal from the Department of Commerce.